\documentclass[nonacm,11pt]{acmart}

\usepackage[colorinlistoftodos]{todonotes}

\usepackage{paralist}
\usepackage{verbatim}
\usepackage{xcolor}
\usepackage{adjustbox}
 
\usepackage{booktabs}
\usepackage{xparse}
\usepackage{color, colortbl}
\usepackage{wrapfig}
\usepackage{tcolorbox}

\hypersetup{
     colorlinks   = true,
     citecolor    = blue,
     linkcolor = blue,
     linkbordercolor= blue,
     urlcolor = blue
}

\NewDocumentCommand{\rot}{O{45} O{1em} m}{\makebox[#2][l]{\rotatebox{#1}{#3}}}%

\newcommand\ggcmtside[1]{\todo[size=\tiny, color=green!40]{GG: #1}}

\newcommand\ignore[1]{}
\newcommand\ASGNMT[1]{}

\usepackage{geometry}
\usepackage{pdfpages}
\usepackage{longtable}
\usepackage{pdfpages}
\usepackage{fancyhdr}
\usepackage{setspace,paralist}
\usepackage{tabularx}

\makeatletter
\def\@copyrightspace{\relax}
\makeatother

\begin{document}

\includepdf{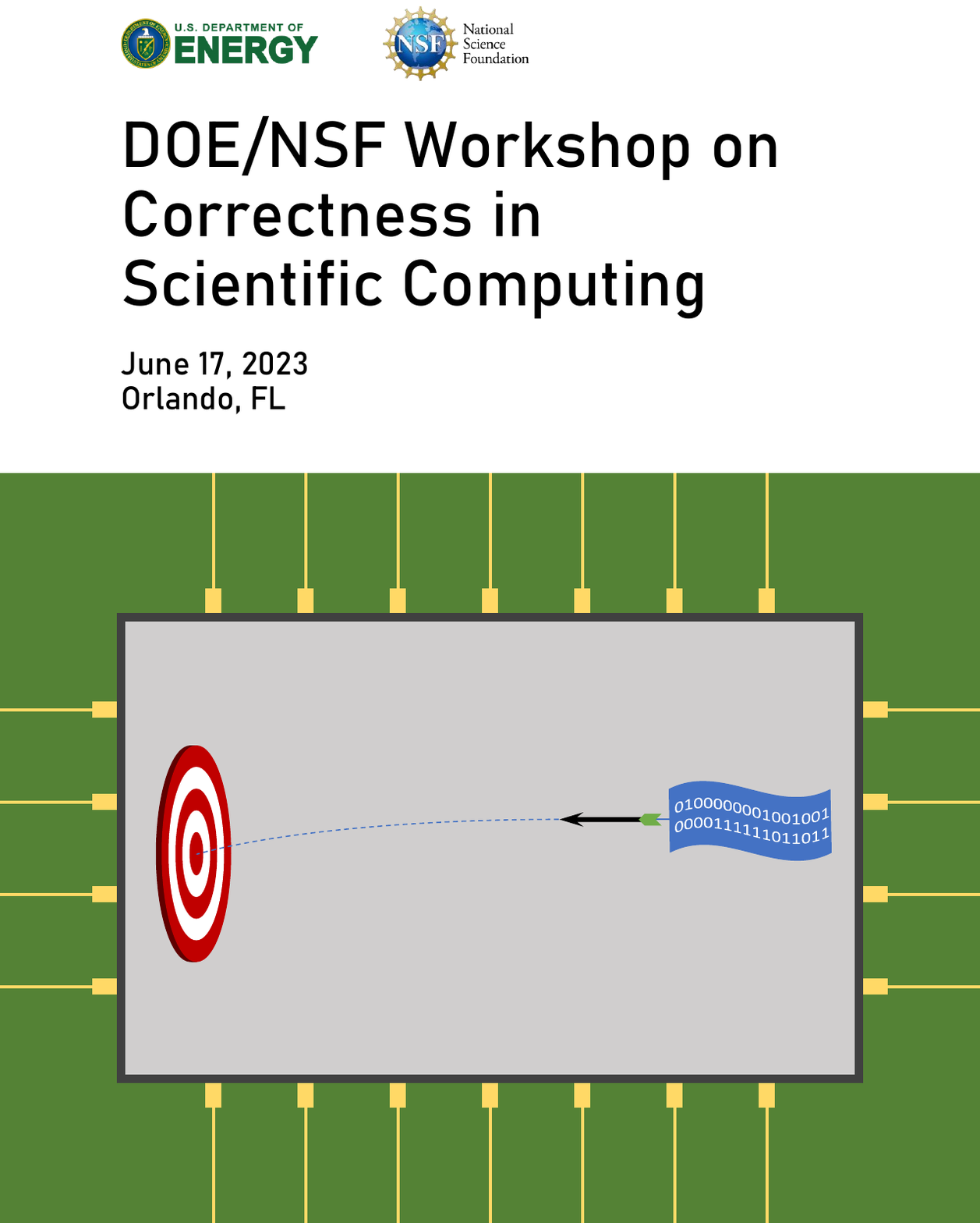}

\singlespacing

\title{Correctness in Scientific Computing}


\author{Maya Gokhale}
\affiliation{%
  \institution{Lawrence Livermore National Laboratory}
  \city{Livermore}
  \state{California}
  \country{USA}}
\email{gokhale2@llnl.gov}

\author{Ganesh Gopalakrishnan}
\affiliation{%
  \institution{Kahlert School of Computing, University of Utah}
  \city{Salt Lake City}
  \country{USA}
}
\email{ganesh@cs.utah.edu}

\author{Jackson Mayo}
\affiliation{%
 \institution{Sandia National Laboratories}
 \city{Livermore}
 \state{CA}
 \country{USA}
}
\email{jmayo@sandia.gov}

\author{Santosh Nagarakatte}
\affiliation{%
 \institution{Department of Computer Science, Rutgers University}
 \city{New Brunswick}
 \state{NJ}
 \country{USA}
}
\email{santosh.nagarakatte@cs.rutgers.edu}

\author{Cindy Rubio-Gonz\'alez}
\affiliation{%
 \institution{Computer Science, University of California, Davis}
 \country{USA}
}
\email{crubio@ucdavis.edu}

\author{Stephen F.\ Siegel}
\affiliation{%
 \institution{Dept.\ Computer \& Information Sciences, University of Delaware}
 \city{Newark}
 \state{DE}
 \country{USA}
}
\email{siegel@udel.edu}

\renewcommand{\shortauthors}{Gokhale, Gopalakrishnan, Mayo, Nagarakatte, Rubio-Gonz\'alez, and Siegel}


\maketitle

 
\tableofcontents

\clearpage

 \addcontentsline{toc}{section}{PREFACE}
 
\begin{center}
    PREFACE\footnotemark
\end{center}
\footnotetext{We acknowledge funding from NSF under the award 
CCF-2319661, CCF-2319662, and CCF-2319663, 
and from the Department of Energy, Office of Science, Office of Advanced Scientific Computing Research under award number DE-SC0024042 and contract numbers DE-AC52-07NA27344 and DE-NA0003525. Any opinions, findings, and conclusions or recommendations expressed in this article are those of the authors and do not necessarily reflect the views of the DOE and NSF.}
This report is a digest of
the DOE/NSF Workshop on Correctness in Scientific 
Computing (CSC'23) held on June 17, 2023, as part of the Federated Computing Research Conference (FCRC) 2023.
CSC was conceived by DOE and NSF to address the growing concerns 
about correctness
among those who employ computational methods to perform large-scale scientific simulations.
HPC systems are beginning to include 
data-driven methods, including machine learning and surrogate models, and their impact on overall HPC system correctness
was also  felt urgent to discuss.
If correctness is not proactively addressed, there is 
the risk of 
producing flawed  
science
on top of 
unacceptable productivity losses faced by computational scientists and engineers.

Stakeholders of correctness were identified to belong to several sub-disciplines of 
computer science; from
 computer architecture researchers who design special-purpose hardware that offers high energy efficiencies;
  numerical algorithm designers who develop efficient computational schemes based on reduced   data movement;
  all the way to researchers in
  programming language and formal methods
   who seek  methodologies for correct compilation and verification.
To include attendees with such a diverse set of backgrounds, CSC was held during the Federated Computing Research Conference (FCRC) 2023.

The program of CSC 
consisted of
keynote talks,
lightning talks, 
and breakout sessions.
The detailed technical program of CSC can be seen at \url{https://pldi23.sigplan.org/home/csc-2023#program}
and is summarized in \S\ref{sec:csc-program-gist}.

\paragraph{Related Workshops:\/}
It may be helpful for the reader to
refer to related past events:
\begin{compactitem}
\item The HPC Correctness Summit held in
2017, resulting in an OSTI report
\url{https://www.osti.gov/biblio/1470989} 

   \item  NSF Workshop on Future Directions for Parallel and Distributed Computing (SPX 2019)
   described at \url{https://sites.google.com/a/cs.stonybrook.edu/spx2019/}.


   \item Workshop on the Science of Scientific-Software Development and Use
sponsored by the U.S. Department of Energy,
Office of Advanced Scientific Computing Research that addressed important sustainability and productivity of HPC software (2021) \url{https://web.cvent.com/event/1b7d7c3a-e9b4-409d-ae2b-284779cfe72f/summary}.

\item The International Workshop on Software Correctness for HPC Applications \url{https://correctness-workshop.github.io/2023/} that was 
held for the seventh time in 2023 at Supercomputing has served to
draw the community together around many important correctness themes.

\end{compactitem}


\vspace{.5ex}
\noindent{\bf Contrast:\/} 
%
Key distinguishing features of CSC are  its more accurate reflection of current realities (e.g., even more heterogeneity and the ascent of ML), given how fast our field is changing.

\vspace{.5ex}
\noindent{\bf Participation:\/}
Our in-person workshop held as part of the Federated Computing Research Conference (FCRC) 2023 attracted a total of 55
experts from academia and laboratories.
%
Most industry invitees could not participate in the workshop--perhaps owing to travel restrictions.  
Nevertheless, the participants were able to voice
many concerns from industry:
a significant number of the participants routinely interact with  industry 
and lend their expertise in arriving at hardware procurement decisions.
A full workshop program of CSC 2023 is
included in the appendices.

%
%

\clearpage
\thispagestyle{empty}

\label{sec:exec}

\begin{tcolorbox}[colback=red!2!white,colframe=red!60!black,title={\centering{KEY TAKEAWAYS}\hfill } 
]

\begin{itemize}
\item Increasing 
hardware 
heterogeneity 
together with partially documented
hardware and libraries
present serious correctness challenges.

\item Code-level
approximations such as
numeric precision selection
can take advantage of  
approximations
already employed 
within the
algorithms being implemented.  


\item  The validity of
performance optimizations in a 
design
must be clearly captured through
suitable correctness arguments.

\item Whenever feasible,
compilers must
 automatically generate  correctness
 checks to validate
 their steps that help aggressively reduce data movement.

\item Suitable ``safety guard-rails'' must be employed whenever AI-generated code is employed. More fundamentally,  today's ML software non-portability requires urgent attention.

\item Checking conservation laws, or
performing ensemble consistency tests
is recommended, especially taking advantage
of reduced-order (surrogate) models.

\item Soft-error mitigation
methods are important to employ,
given the increasing transistor counts, interconnect complexity, and the use of
aggressive energy reduction
techniques.



\item Focused efforts to
demonstrate end-to-end correctness through case
studies is highly desirable as it
can help galvanize players across the abstraction stack.

\item All said, we need more tools, including to check for basic  properties such as reproducibility and determinism.

\end{itemize}

\end{tcolorbox}

\section{Executive Summary, Findings}

Continued improvements in
High Performance Computing (HPC)
are essential to sustain progress in  research and engineering practices in an ever-growing
list of scientific computing
areas including precision
manufacturing, simulation of modern power grids, fighting emerging pandemics, and climate
modeling.
%
In the modern context, HPC-based
simulations are
carried out under tight time schedules, running highly optimized applications on cutting-edge hardware to obtain the shortest possible
time to solution. 
Unfortunately, at the bleeding edge
where hardware performance improvements are slowing down,  
there will be increased
use of more aggressive algorithmic design approaches
involving reduced precision arithmetic 
and various approximate computing methods, all of which will make correctness even harder to achieve (and even define).
%
The supporting hardware and software  will become even more heterogeneous, employing
multiple concurrency models, different numerical schemes,  different number representations, and a variety of accelerators including GPUs, Tensor Cores and Matrix Accelerators that differ from each other in subtle ways.
At this pace and scale, errors can occur at all levels---while developing new applications and porting existing ones---risking dissemination of incorrect results, release of
flawed products and ultimately slowing down scientific findings and discoveries. 

 Despite these serious consequences, 
the levels of
 correctness
 needed to help
 reduce these risks
 are not being targeted.
A variety of factors
may be attributed for this 
lack:
 (1) tight project
deadlines that may force decisions which (in hindsight) prove
to have detrimental correctness consequences,
(2) focus on achieving maximal performance,
which may encourage exploiting lower levels of APIs
(e.g., vendor-specific synchronization methods)
which prove to be difficult (or error-prone) to port,
but most importantly (3) lack of exposure to tools for formal reasoning
about the correctness, to the extent that
even using today’s well-established light-weight methods
are overlooked.

We advocate elevating correctness as a fundamental software
design requirement, spanning low level libraries through complex multi-physics simulations and emerging scientific workflows.
%
The 2023 workshop on Correctness in Scientific Computing (CSC'23) was organized to re-examine
correctness notions basic to improving the state of knowledge as well as practice in this area.
The remainder of this executive summary---deliberately kept succinct and high-level---helps focus on 
the key takeaways from this workshop.
%
%
Section~\ref{sec:intro} provides basic definitions 
and area-specific details, pointing out what was omitted from this report.
Beginning with Section~\ref{sec:challenges}, 
we list
key areas to target,
key techniques to further develop, and point out
specific tools and intervention opportunities.
We close with a  short list of 
challenge problems 
and appendices listing details of
the workshop participation.

\paragraph{Correctness of Scientific Computing Systems:\/} 
While the notion of correctness in scientific computing is many-faceted, it is still important to state the general principles of correctness as well 
as example violations thereof.
At the level of hardware and software, correctness can be understood with respect to the services provided,
including avoidance of excessive numerical rounding, floating-point exceptions, 
data races, deadlocks, 
and module interface assertion violations. 
There are of course difficulties when attempting to refine these notions. 
For instance, the definition of ``excessive'' numerical rounding error is highly problem-dependent; and a vast majority of HPC modules in use today do not carry interface assertions regarding numerical stability.
However, even at infinite precision, models
could be wrong, being an approximation of reality---requiring strong validation 
methods in addition to verification.

We may also distinguish between the
correctness of 
the code and correctness
of the overall 
scientific application.
For example,
one standard approach that helps show code correctness is by formally arguing
that program loops maintain their
loop invariants and also terminate. 
A similar necessary step in many applications
is when the scientific process being simulated (e.g., a chemical reaction) quiesces, while respecting mass and energy conservation laws. 

Having well-defined correctness criteria at both levels 
gives the community 
the opportunity
to define
specific instantiations of these ideas in their domain, and build tools that further elaborate as well as support correctness checks.
A good set of definitions
also helps identify what is
lacking in terms of
pedagogy.
One measure of success in these endeavors will be reaching a point where the
HPC community treats correctness as a basic necessity.
%


\paragraph{End-to-end Correctness:\/} 
As HPC involves layers of abstractions beginning with the targeted problem and how it is mathematically modeled---all the way to specific CPU and GPU code, 
aiming for end-to-end correctness is
crucially important to close potential
gaps in the correctness stack.
Tractable end-to-end  correctness 
approaches are also ideally layered,
matching the abstraction layers.
This report specifically focuses on
the lower levels of this abstraction
stack, given that this is currently
facing the brunt
of the stress from the ending of Moore's
law (discussed momentarily).
Based on this focus, end-to-end 
correctness must
prioritize creating rigorous
models for the computing
hardware, and
then showing that
the software running with respect to this hardware
model behaves as required.
Clearly, one has to also close gaps
with respect to the higher levels
of intent, of the
modeling process,
and validate the assumptions that
connect adjacent layers.
The overall approach
taken to achieve end-to-end correctness
will typically involve the use of
uncertainty quantification
(e.g., to account for noisy data),
checks of
agreement with real-world observations
(e.g., pointwise validation against real-world
measurements),  
and statistical methods (e.g., similar to those used to
provide projections for hurricane trajectories).

%

%

\begin{figure}[ht]
\vspace{1ex}
\begin{minipage}{0.46\textwidth} 
  \centering
    \centering
\includegraphics[scale=.1]{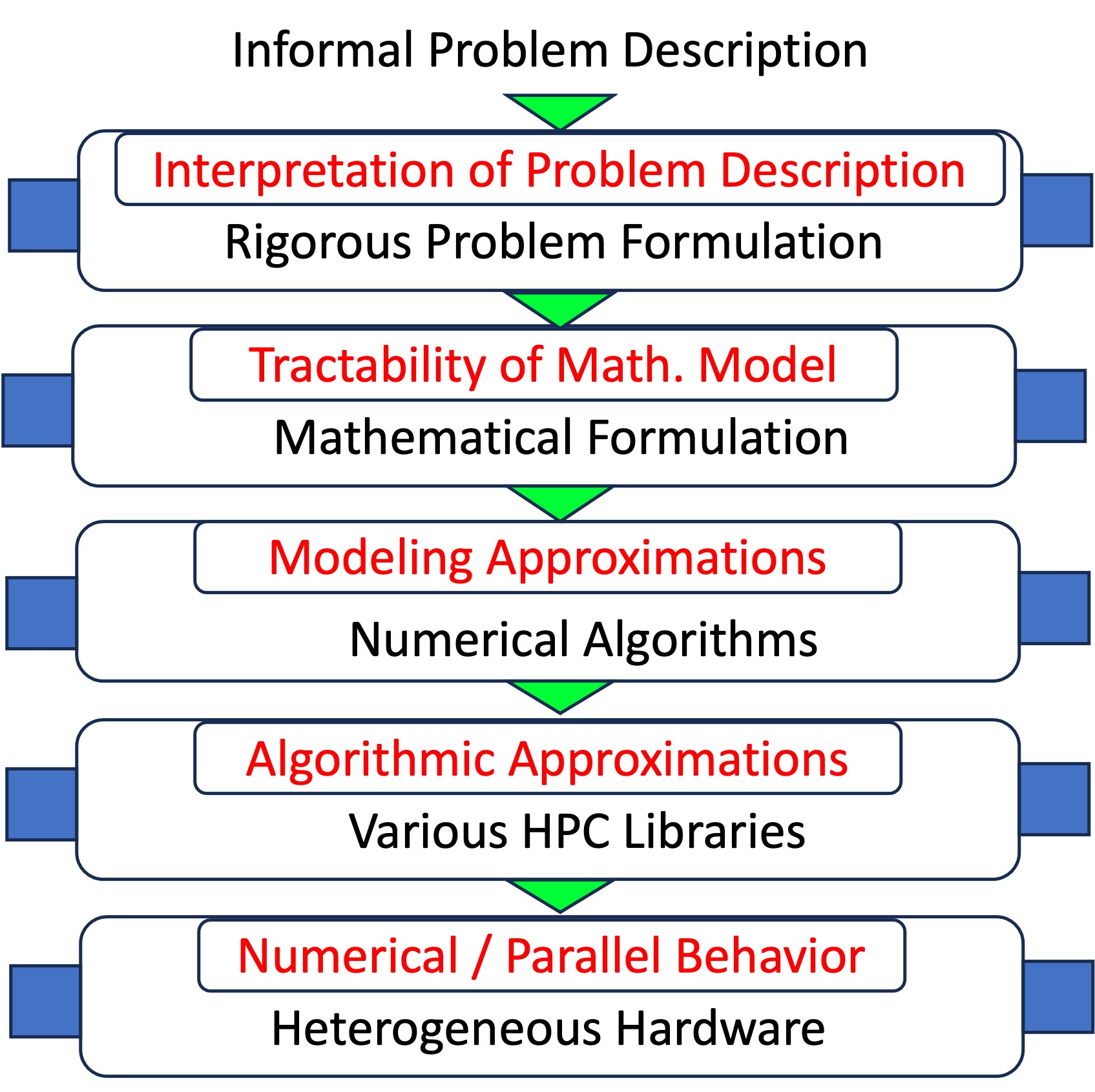}
    \caption{End-to-end Correctness in HPC. Each item in black (e.g., Rigorous Problem Formulation) is subject to an implementation consideration or approximation (e.g., Tractability of the Mathematical Model in red color) to be more concretely realized at the next lower level of abstraction (e.g., Mathematical Formulation).}
    \label{fig:end-to-end-correctness}
\end{minipage}%
\hspace{1em}
\begin{minipage}{0.46\textwidth} 
\vspace{1ex}
 \begin{tcolorbox}
[colback=red!2!white,colframe=red!60!black,title={\centering{Case Studies at Various Scales}\hfill } 
] 

\begin{itemize}
\item  
A small-scale 
study might be to take a specific conjugate-gradient solver and argue its correctness when deployed on a new CPU/GPU platform using a specific open-source compiler and its optimization flags.

\item 
A   medium scale  study 
might be the end-to-end reliability of a finite-element package when  ported  from  CPUs to  GPUs.

\item 
A longer-term study
might focus on  the analysis of
electrical power distribution networks.
Correctness properties verified
might include grid stability,
and also   
the software stack,
including whether
the numerical solvers
can handle real-world
problems~\cite{kasia-solvers}.
 
\end{itemize}
\end{tcolorbox}
\end{minipage}
\vspace{1ex}
\end{figure}


%

Achieving end-to-end correctness in HPC involves closing various gaps (shown in red in Figure~\ref{fig:end-to-end-correctness}): in formulating the problem, choosing a suitable mathematical model and numerical algorithms, binding it to specific HPC libraries and dealing with their numerical behaviors.
Case studies demonstrating  end-to-end correctness
can help
close these gaps,
 and hence 
 are highly
 recommended
to be pursued.

They must be chosen carefully, conducted by involving a sufficient number of
stakeholders and know-how providers, and scheduled to finish in a 
timely manner to remain relevant.
It might be most beneficial to aim for studies
that finish within a year (or  
have clearly defined yearly deliverables) {\em thus 
permitting these studies to 
be discussed
at relevant annual workshops.}\footnotemark
%
%
\footnotetext{One such initiative has already taken root---namely the ``HPC Bugs Fest'' organized as part of the 
International Workshop on Software Correctness for HPC  Applications~\cite{correctness2023}.}
%
%
%

Many of the aforementioned
case studies
would be impossible
to conduct 
{\em without
help from commercial
firms}.
This
is because
the vast majority of today's
HPC systems are built around
closed-source vendor-provided libraries and
computing hardware
that carry scanty specifications.
For example, although
many of today's GPUs have
assembly-language instructions sets, these ISAs do not have the same degree of standardization that traditional CPU ISAs have enjoyed, as evidenced by numerous reverse-engineering attempts at discovering their properties~\cite{jia2018dissecting,fuzzing-ptx,ptx-math,hpdc2023-gpu-fpx}.
%
These problems are likely to get worse when processors 
designed by ``hyperscalers'' such as
AWS, Microsoft, Google, and Meta become attractive options due to their cost/performance advantages.
How the open-software community can keep up with the challenge
of discovering adequate specifications for these components remains an unsolved challenge.

The {\em opportunity costs} of scientific computing software bugs
can be arbitrarily high.
If a bug surfaces during an experiment that cannot be
repeated (e.g., processing a rare natural phenomenon), the world
community stands to lose.
This is one reason why we absolutely must have hardened and trustworthy
components at our disposal.
Unfortunately, practical experience (e.g., inability to handle
``difficult matrices~\cite{kasia-solvers}'') suggests that
we are really not prepared in this manner.



\paragraph{Performance/Correctness Entanglement as Moore's Law Wanes:\/}
Thanks to the multiple decades of 
research in computer science on the principles of correctness checking,
there are, {\em in principle}, 
well-understood techniques that can help us establish end-to-end correctness in scientific computing systems. 
For instance, synthesis techniques to generate arithmetic libraries that perform correct IEEE-compatible rounding exist~\cite{lim:rlibmall:popl:2022}; formal techniques to verify numerical schemes to have the desired correctness properties exist~\cite{tekriwal-etal:2023:jacobi,boldo-melquiond-book}; and mechanically verified compilers also exist for simpler dialects of languages such as C~\cite{leroy2004compcert,xavier-fp}.
While some simple codes can be argued end-to-end correct using 
these techniques,
tackling the correctness of
larger codes requires significantly
more concerted effort.

There is now a
clear uptick in the level and variety of
correctness 
challenges  faced.
This is caused by
the barrage of emerging concerns originating with the end of Moore's law---specifically, the end of Dennard scaling.
In practical terms,
this translates into 
the inability of conventional computing capabilities to keep pace with the scale of problems to be solved.\footnotemark
\footnotetext{Dennard's scaling~\cite{dennards-scaling}, when it held
before  year 2006, allowed us 
to, in principle, put
more transistors~\cite{spectrum-transistors-75} 
on a chip without leading to
meltdown, thanks to commensurate voltage-reductions
that were possible---but no longer.}
%
%
What this means
is that near-term performance boosts are likely to mostly
come from the combined use of multiple types of CPUs and accelerators (in particular GPUs), multiple memory systems, 
and novel types of interconnect technologies.
%
%
%
%
%
Highly tuned accelerator
programs involve
memory barriers, fences, and other
synchronization mechanisms,
the manual introduction of 
which is error-prone.
%
%
%
Vendor documentation in this space is often 
lacking or unclear, and there are
no systematic cross-platform porting approaches
that help migrate the synchronization primitive 
employed.
%
Violated concurrency discipline can turn into
grossly miscomputed results, deadlocks, or   {\em data races}---essentially uncoordinated memory accesses, but 
also an eventuality
that breaks the assumptions under which many compiler optimizations are made.

The topic of data races 
provides a good example
of the ``last-mile neglect''
that seems pervasive in HPC-correctness
as a whole, which slows down technology transition into practice.
%
Even with all the technology at hand, data race checking took some time before it was offered in a highly usable form as ``sanitizer'' flags in {\sc{gcc}} and {\sc{clang}} (largely through Google's  efforts).
The same technology took a while before it was engineered into the {\sc Go} language ecosystem (again by Google).
Similarly today, despite all the know-how that exists, there isn't a widely known and effective data race checker for GPU programs: the best of company tools do not check for global memory races, rendering them incapable of checking races
in a significant 
fraction of programs in use.
In the last 15 or so years, many academic tools were contributed to handle GPU race-checking~\cite{DONALDSON20173}, but unfortunately lacking industrial backing and engineering effort to maintain an adequate front-end, 
ceased to be usable in practice.
%
Tool extensibility is also
quite important, given that 
new synchronization primitives
will be added with each new
GPU generation.

%
%
%
In addition to uncaught data races that result in non-reproducible code behavior,
cross-platform portability 
as well as trust in compilers
has steadily diminished 
with the increasing use
of hardware ``meant for other other purposes'' (e.g., ML)
which HPC practitioners
are incentivized to use~\cite{cloudy-uncertain-reed}.
It has been observed in an ongoing
study that
the floating-point details  of GPUs---especially   Tensor Cores---can vary in subtle ways with respect to their rounding behaviors, which can hinder
direct cross-platform
software porting with assurance.
%
%
%
 %
%
Compiler flags 
continue to have
less known properties, and as
reported in one study 
caused scientists to spend an extra month trying to root-cause an issue in a climate simulation program called CESM~\cite{science-on-keel-cacm} (finally attributed to
a NaN or ``Not a Number'').
In another study~\cite{flit-hpdc19}, it was observed
that an IBM compiler
evaluated a C++-undefined construction
into a NaN (``Not a Number'') while
a GNU compiler did not, resulting
in non-reproducible test outcomes.
In a third study, it was
observed that NVIDIA's compilers
may generate FP32 instructions that bind to special function units when the {\tt fast-math} optimization is invoked, when without this flag, the code may have been of FP64 precision~\cite{hpdc2023-gpu-fpx}.
The programming language
community has not spent time 
characterizing formal semantics 
to this level of detail.

%

In the past, gaps in our understanding were
often surfaced through
``constructive attacks'' 
in well-cited academic publications.
Notable examples of this kind  
include formalization of the x86 memory consistency model~\cite{x86-sewell}, Java Memory Model characterization~\cite{pugh-java-studies}, study of GPU weak-memory behaviors~\cite{alglave-asplos15}, and studies
of LLVM optimization pitfalls~\cite{csmith}. 
%
%
It is clear that in this period
where Moore's law 
is ending and
HPC is forced to adopt ML-specific
hardware 
that is designed to be more energy-efficient
and performant,
vendors
can afford to do only a limited
amount of specification generation before they release a new component.
One would thus hope for 
a new world order
where many of the aforementioned constructive attacks
occur within studies sponsored by the vendors, resulting in much better (and timely arrival) of specifications.\footnote{Perhaps with the added benefit of the hoopla avoided.}

\paragraph{Making HPC-algorithm-specific Approximations and Abstractions:\/} The success of
scientific computing fundamentally rests on various abstractions and approximations: very few problems can be solved as stated, without employing a sufficient number of them.  
As a concrete example,  
multigrid methods (employed to accelerate
the convergence
of numerical solutions)
are based on abstractions and approximations that coarsen the grid before 
solving---offering a thousandfold increase in convergence rates.

Given this context, 
it is important to ensure that newer approximations be compliant with the nature of the HPC-algorithms in use.
For example, in their quest to reduce data movement~\cite{data-movement-springer,data-movement-shalf}, programmers often entertain additional  approximations 
that help reduce the volume of
data moved---lowering the numerical precision and employing lossy
data compression being two popular choices.
A natural question to ask in this setting is
whether these approximations are compatible with those already made in modeling the problem as well as made in the HPC algorithms themselves (e.g., in~\cite{mccormick-mg-precision} a rigorous analysis of mixed-precision multigrid methods along these lines is presented).
Ideally, ``compatible'' must mean {\sl well below}---meaning, the numerical errors must ideally be a small fraction of the modeling error. 
While this fact is somewhat well-known, {\em studies/tools to check whether such relationships hold} are lacking---and hence, highly recommended.

A higher-level takeaway
is that the specification of a scientific computing component or application, however it is realized, must be expressed using appropriate domain abstractions.  The abstractions are not only about numerical accuracy---although that is part of it.  They are needed to make specifications understandable and succinct; they should reflect the way the scientist actually thinks about the algorithm.  Likewise, appropriate verification techniques must be designed around those abstractions.

The space of ``compatible approximations/abstractions'' can be
a long-term focal point for the correctness community.
{\em This discussion must also include the long-term goal of developing algorithms that are tolerant to errors and faults.}

\paragraph{When Implementations become Specifications: Differential/Metamorphic Testing:\/} Whenever artifacts with two different realizations are available, each can act as a specification for the other.
In HPC, one has access to such pairs of comparable artifacts in many areas:  MPI libraries, OpenMP libraries, GPUs, arithmetic libraries, and data compressors for instance.
This offers a rich set of possibilities for differential testing, i.e., checking one type of realization against the other (e.g, differential testing of numerical libraries~\cite{DBLP:conf/issta/VanoverDR20}). 
Metamorphic testing further aims to introduce changes that ought not matter at a certain level of abstraction (e.g., $f^{-1}(f(x))=x$) and observes whether such invariants are violated by transformations (e.g., by a compiler).
This rich set of possibilities is yet to be fully exploited in scientific computing.
Traditional correctness research demonstrates the huge returns on investment that are possible through differential verification:  
the so-called ``semantic fusion''~\cite{semantic-fusion} work
 has pitted the CVC4 SMT-solver against the
 Z3 solver, exposing thousands of bugs.

 Added benefits are obtainable
if random program/input generation  (pioneered 
in works such as CSmith~\cite{csmith})
are employed.
While random program generation similar to CSmith
exists in scientific computing (e.g., Varity~\cite{varity}), many more such possibilities remain unexploited in HPC.
One deficiency of random program generation is that even such approaches must be heuristic-driven to suitably contain possible exponential case-explosion.
A complementary approach might be to also extract numerically ill-conditioned kernels from existing HPC programs and turn them into self-contained test programs.

\paragraph{Opportunities and Challenges of Combining HPC and ML:\/} The nature of scientific computing and HPC
itself is changing.
It is now
 enjoying growing support from data-driven approaches based on machine learning (ML), including the use of  surrogate models and physics-informed neural networks.
 The use of AI-generated software in scientific computing is also expected to grow---but again with very few correctness checks in place as things stand today.

While HPC adjusts itself to this use of AI,   
many synergistic opportunities
seem to open up.
For example, one hopes that everyone writes assertions at the software module boundary level.
Such assertions are, unfortunately, rare to find in practice.
While it is possible to write code-level assertions (e.g., capturing the journey across the state space of each prominent variable through pre- and post-conditions), 
one has to question whether such attempts truly capture the {\sl essence of the computation}---the physics or chemistry being modeled---or only the ``inner vibrations'' of the code.\footnotemark%
\footnotetext{Akin to, say, viewing music as pressure waveforms. Our point is that the modeling and the physics are centrally important in HPC, and the specific code that is executed need only ``answer to'' the physics.}
The point being made is that while the integrity of the actual code that implements an HPC algorithm will likely be in the traditional 
precondition/postcondition/invariant style, 
the {\sl intent} or {\sl expected higher level behavior} of a code unit may be to implement a piece of functionality that may well be captured by 
conservation and symmetry laws of Physics---or by actual AI-based surrogate models. 
In this context, it seems attractive to consider some of these surrogates serving the role of assertions.



A key research angle can be the ability to effect tradeoffs between hardware/software correctness and physics-level correctness, 
mediated by abstractions/approximations that help meet the physics goals despite sacrificing some of the tight guarantees at the code level.
As an example, one can imagine rewriting the body of a loop with another that, despite generating a higher overall numerical error has other more desirable attributes (such as
helping eliminate floating-point exceptions, offering more reliable loop-convergence guarantees, and/or higher performance).
The recent rise of {\em symbolic regression}~\cite{symbolic-regression-contest} which synthesizes program fragments to fit the observed data seems to be such a ``physics-first'' approach to arrive at working code. 
It seems worthwhile to study how such code rewrites compare with rule-based rewrites such as in tools like Herbie~\cite{herbie}.


\paragraph{Correctness Notions around Data Movement Reduction:\/}
Data movement reduction, an important prerequisite to achieving high performance, is often achieved
through  compiler-based locality enhancing optimizations.
While the algorithmic correctness of these optimizations
rests on dependence-preserving transformations,
they have often proved to be buggy in terms of how they are implemented as optimized programs.  
This area needs constant attention due to the continued arrival of newer transformations, their complexity
especially at the implementation level, and subtle interactions among multiple transformations.
Fortunately, these transformations are often part of automated frameworks such as
TACO~\cite{taco} and TVM~\cite{tvm}, thus offering the possibility of verifying the code of transformation frameworks rather than individual transformations. 

\begin{wrapfigure}{r}{.6\textwidth}
\input{sections/callout-bit-flips.tex}
\end{wrapfigure}

An orthogonal way to achieve reduced data movement is 
by reducing the size of the data moved using lower precision number representations.
This approach is widespread in machine learning because of two reasons: (1)~the process of loss minimization through stochastic gradient descent largely cares about the  dominant
directions
 of the derivatives---and not the error in each derivative; (2)~dynamic rescaling of values and gradients (so that they do not exceed the allowed dynamic range of floating-point values which risks creating erroneous NaN-values) can be more readily employed. 
HPC implementations do not readily lend themselves to tolerating errors in this manner.


Unhandled out-of-range values in HPC can cause floating-point exceptions, the most egregious of them being NaN (not a number) as well as overflow/infinity
values. 
These exceptions can ruin the overall logic of the application~\cite{DBLP:conf/correctness/DemmelDGHLLLPRR22}, and additionally, today's dominant GPU types do not trap such exceptions in hardware---in principle demanding that every runtime value be checked and ensured to be not an exceptional quantity such as a NaN.
As things stand, the community seems
unaware of the extent of this problem---and those who are aware have no means of ensuring safety under all circumstances.

For instance, if the deterministic ordering of addition is not guaranteed
when summing
an array of numbers, overflows might occur under some of the orderings.
While the costs of enforcing determinism
can be prohibitive, facilities to reinstate determinism in order to facilitate testing are highly valuable.


Another important avenue to achieve
significant levels
(e.g., by a factor of 50$\times$ or more in some studies) of data movement reduction
is to compress large data items
before transmitting them.
In most cases,\footnotemark the data is decompressed 
before being put to use. 
\footnotetext{Unless direct operations on compressed data (without decompression) are supported---as are happening in the evolving field of {\em homomorphic schemes} on compressed data (e.g., \cite{martel-blaz}).}
Unfortunately,
the decompression and compression steps 
introduce errors whose proportions are known to some extent---but not sufficiently well enough to account for all their correctness consequences (e.g., will such errors accumulate detrimentally?). 
Closing the correctness gaps in this area in as many specific domains as possible is a highly recommended direction for further research. 
There are  rigorous analyses that point to possible approaches~\cite{fox-iterative-schemes}.

\paragraph{The Impact of Soft Errors:\/}
As the impact of extreme heterogeneity ripples through the design stack, and as we push toward the highest achievable performance in very complex chips that contain 100s of billions of transistors,
low-level effects that cause transistors to misbehave are a growing concern.
A considerable amount of research has occurred 
under the banner of {\em system resilience}~\cite{resilience}. 
There has, unfortunately, been very little adoption in actual HPC system designs and codes---primarily due to the unacceptable overheads of today's solutions, largely
attributable to
the non-trivial costs of redundant computations involved.

Now, however, as we approach the 1 nanometer manufacturing node (expected in a year or two), with system complexity driven by chiplet-based design, 3D stacking, and the systems-on-chips (SoC) model---resilience is reappearing on the priority list of the computing community.
%
%
In this regime, the number of soft errors that can be adequately guarded through error detection and correction techniques is as yet unknown.
Jumps in transistor density and the vastly increasing number of interfaces, interconnects and packaging options offered by chiplet-based designs and SoC interconnect protocols have as-yet unknown reliability ramifications.
While studies of the effectiveness
of cross-layer resilience~\cite{bosedac19}
provide many useful directions to pursue, 
the proposed solutions clearly
need to keep up with
the growing complexity of
systems.

These hardware ``bit-flips'' can be surmised to occur based on the sheer size of digital systems today---and yet, their effects are very difficult to observe at system interfaces.
Additional considerations such as {\em wearout} (the fact that transistors ``age'' and drift in behavior over time) must also be kept in mind.
Even though hardware component manufacturers either provide or are working on error detection/correction logic strategically at various interfaces (e.g., memory interfaces), not every stage (especially, compute-intensive stages) can be protected in this manner without incurring severe slowdown.
This implies that  HPC and ML programmers must employ additional logic to take into account bit-flips that are not handled at the hardware level.\footnotemark
Approaches to
develop detailed solutions
along these lines are a 
clear research priority for
our area.

\footnotetext{In some HPC codes
and many more ML codes,
programmers may be able to ignore certain errors,
given the inherently error-tolerant nature of many of their applications.
This is reflected in some efforts~\cite{bose-low-voltage-training} where researchers
have proposed approaches
that exploit hardware {\em non-resilience} induced in a well-controlled manner through voltage reductions, resulting in significant energy savings. Prudence is well-advised before these methods are widely adopted and trusted.}
%
%


In summary, 
achieving
correctness in scientific computing crucially relies on personnel (with skills in specific correctness methods) and practices (that are 
well-designed and widely adopted), and must be supported by scale-appropriate 
checking tools.
The tools must be incisive (detect
bugs closer to the root-cause) as well as informative (produce sufficient debugging information).

\label{page-of-challenges}
\begin{tcolorbox}[colback=red!2!white,colframe=red!60!black,title={\centering{Assortment of Challenge Problems}\hfill } 
]
\begin{itemize}

\item Create mathematical libraries that carry rigorous semantics that are preserved under porting to new platforms and/or subject to optimizations by different compilers.
E.g., if $sin$ and $cos$ use Taylor approximations (or numerical approximations of ULP error) resulting in $sin_1$ and $cos_1$,  then show that
$sin$ matches $sin_1$ and and $cos$  matches $cos_1$ in a precisely defined way.

\item Develop systematic methods
for implementing, testing and verifying numerical libraries on a variety of heterogeneous architectures.

\item Investigate
the use of 
permission-based separation logic for reasoning about GPU programs
(see~\cite{huisman-vercors}),
targeting GPU programs from popular textbooks.   

    \item Develop cutoff bounds for various concurrency problems.
The so-called ``two-thread theorem'' for GPU symbolic verification is one such bound~\cite{DONALDSON20173}.

\item Develop
methods for adapting existing (verified) libraries to new situations (new precision, new memory layouts).

\item Arrive at specifications, and then 
verify (parts of) established
libraries such as
PETSc, MPI libraries, and BLAS.

\item Verify specific properties about established 
OpenMP runtimes.

\item Write the specification for 
a simple multigrid solver. Formally characterize what the ``V-cycle'' 
achieves. Following McCormick's work
at~\cite{mccormick-mg-precision} may be helpful.

\item Attempt to formalize one of the
Machine Learning models given as a benchmark in a compendium of ML and surrogate models coming from multiple domains~\cite{AI-Benchmarking-For-Science}.

\item Write an end-to-end formal specification for a simple lossy data compression scheme such as PyBlaz~\cite{drbsd} which supports homomorphic operations in the compressed space.

\item Verify programs that obtain low rank approximations for matrices and tensors.

\end{itemize}

\end{tcolorbox}

 \clearpage




\section{Basic Definitions, Urgency of Problem, and Report Scope}
\label{sec:intro}
 
\subsubsection*{What does correctness mean?} 
%
In HPC, one can divide the definition into {\em program-level correctness} and {\em application-level correctness}.
At the program level, the correctness of a code module entails, among other things, these: (1)~elimination of data races or deadlocks,
(2)~handling of floating-point exceptions, (2)~conforming to acceptable floating-point approximation bounds, (3)~ensuring that compiler optimizations applied to the module respect source language semantics, and (4)~the input/output behavior obeys user-specified assertions.
%
%

At the HPC application level, correctness
of a 
scientific computing system
entails, among other things: (1)~fulfillment of 
the scientific
mission (which in turn requires
the ability to carry out
long-running simulations), (2)~reproducibility
when handed over to another group or ported to another platform, and
(3)~consistency and repeatability of
behaviors at very high scales (e.g., when replicated over many more threads)
%
with respect to those that are more stringently (and affordably) checked at lower scales.
%

Program-level correctness might adopt 
various program logics, abstract interpretation and model-checking
 to determine whether (and when) an application-level simulation maintains its vital invariants
 during execution
 and also terminates.
 %
%
Additional checks such as for deadlocks and floating-point exceptions (also innately program-level checks) would also be employed as needed. 
%
HPC application-level correctness 
is, in addition, often directly tied
to the behavior of the scientific
phenomenon being simulated.
For instance, an HPC application simulating a chemical reaction terminates when the reaction itself
attains quiescence~\cite{runtime-prediction-hpc}.
Application-level correctness  in the context of climate simulations (as an example) might also involve tests such as {\em ensemble consistency tests} which ensure that the principal components of observed climatic patterns fall within acceptable bounds~\cite{baker-ensemble-consistency}.
%

Application-level correctness strongly depends on how the science gets modeled based on modeling assumptions (e.g., spatial/temporal discretization resolutions) and abstractions (e.g., ignoring gravitational influences from distant bodies in N-body simulations).
While the two correctness notions discussed are best kept separate, they can get entangled, for instance, if uncaught program-level errors
invalidate application-level abstractions and approximations.
Exploring these entanglements and ameliorating their detrimental effects is a very worthwhile research direction.
The following are three of the many such intervention opportunities that exist:

%

\begin{compactenum}
    \item One must ensure that lowering the precision of the data types in a  program loop does not 
    significantly alter
    its termination characteristics
    (as has been observed~\cite{ichi-gmres}).
    Efficient methods for
    (re-) verifying the loop termination criteria
    across whole data ranges
    are essential.

\item While a surrogate model
built using one of the physics-informed neural network varieties (PINN~\cite{Karniadakis2021,li2021fourier}) may offer significant advantages (e.g., higher simulation speed, better fit with observed data) over existing conventional-style HPC codes,  caution in using these newer ``mixed-modality simulations'' (some HPC and some ML)  must be exercised, as with any new  technology.
For example,
as with neural networks used in
 other applications, if the underlying 
 neural networks are sparsified to gain
 higher efficiency~\cite{han2023retrospective}, how can one be sure that there are no semantic
 misclassifications introduced~\cite{cie-hooker,joseph2021going}?

\item When using data compressors,
it must be ensured that
the error (or bias) introduced by the decompression/compression pipeline does not contain artifacts---such as periodic noise patterns corresponding to the fixed-sized blocks that the compressor used inside~\cite{fox-blocking}.
Such noise patterns can invalidate the assumptions made by the HPC/ML processes using the decompressed data. 
Ways to incorporate program verification methods (based on pre/post conditions) will be interesting to develop in this setting.

\end{compactenum}
%

%

%

%


\begin{wrapfigure}{r}{.6\textwidth}
\input{sections/callout-climate-correctness.tex}
\end{wrapfigure}

\paragraph{Why emphasize correctness now?}
 Correctness research in computer science 
 was pioneered by 
 Floyd~\cite{floyd1967},
 Dijkstra~\cite{dijkstra}, Hoare~\cite{hoare:1969:axiomatic},  
with selected exemplar
results in recent decades being 
 the original 
Astree system~\cite{astree}
(now being developed by AbsInt~\cite{absint}) 
and the CompCert~\cite{leroy2004compcert} compiler effort.
These are examples
of non-trivial systems 
where
correctness guarantees
approaching the ideal
of {\em correct by construction
for all inputs}
is achieved.
These are also examples 
that demonstrate the need to 
nurture 
  early pioneer visions  
  over  decades
before highly reliable and practical solutions emerge.
Thanks to similar results,
Intel~\cite{roope-intel-fv-i7}
had once managed to entirely dispense
with traditional simulations in
the verification of arithmetic
units.

Along these lines, the time is now
ripe for ``self reflection''
by the computing community:
{\em where we would end up if the status quo prevails?}
Hopefully, decisive steps to correct the course will be taken based on planned moves rather than forced moves due to dramatic software failures.
 For instance,
we are now in an era of fast-paced changes
where even if we are given the full source code and the complete hardware schematic of an installation, one cannot predict the behavior of 
the software---either {\em barely}, or 
{\em only as an ephemeral success}.
We now elaborate on this point.

The latter case of ``ephemeral success'' is easier to explain.
It is well known that it takes time and effort
to build up correctness machinery and tooling in support
of emerging number systems, newer 
concurrency paradigms, etc---and these emerging options
are changing quite frequently in this era of high
heterogeneity in computing.
Tools that are too closely tied to specific 
choices in this space only provide ephemeral
value.
On the other hand,
tools that {\em anticipate} future trends
and are designed to be  parametric
 (parametric with respect to memory models,
 number system choices, etc.) 
 can provide longer-term value.
 Overall,
 such tools are to be preferred 
 if the benefits of long-term value
 far outweigh the burden of 
 accommodating parametricity.
 

The ``barely'' cases are much more worrying---and typically caused
by manufacturers {\em not revealing details of components}.
Examples in
hardware include~\cite{demystifying-ampere,demystifying-cuda-warp-divergence,alglave-asplos15,non-monotonic-fassi-higham,rising-heterogeneity}
and examples in critical lower-level software include~\cite{ptx-math}.
This leads to
reverse-engineering attempts~\cite{non-monotonic-fassi-higham,fuzzing-ptx}
that might help make some progress~\cite{ptx-math,hpdc2023-gpu-fpx}---but, ultimately,
such ``experimental discovery'' of specifications is bound to be suspect.
%
%
Others have discovered ``natural'' properties (e.g., monotonicity during addition\footnotemark)
\footnotetext{The result of adding a set of floating-point numbers
increases if any element is increased.} 
do not hold~\cite{mikaitis2023monotonicity}---``{\sl is a feature that may have appeared unintentionally as a consequence of design choices that reduce circuit area and other metrics.}''
%

%
Persistence on the part of the academics around central APIs such as
the x86 instruction set architecture and 
LLVM have led to better specifications~\cite{x86-sewell} and 
tool-chains~\cite{alive2}, respectively.
But there are too many such APIs and libraries---implying that 
industry has to step forward and engage as partners in creating
well-understood specifications.
An added advantage for such industries would be 
their ability to
maintain 
proprietary advantages while helping enhance the reliability and
usability of their products.

We are entering a new era 
of programming
where automated general-purpose code generation has become more accepted, thanks to systems such as the Github Copilot. 
Automated code generation is being
viewed with great interest also in HPC
due to the allure of increased
programming productivity.
Due caution must, however, be exercised
before whole-heartedly embracing this technology in HPC
for several reasons, some being
these: (1)~subtle flaws (e.g., an accidental negation) may be hard to notice in auto-generated code; (2)~aberrant behaviors in HPC 
simulations 
may take much longer to manifest (e.g., many thousands of simulation
steps) and may need ``expert eyes'' to  detect them ; and (3)~reliable automatic code
generators will likely  
have to be 
driven by clear and air-tight
formal
specifications that capture the 
behavior of the code to be
generated, and writing such
specifications may be harder than
writing the code itself
(see the proof effort reported
in~\cite{boldo2013wave} for a ``simple'' wave equation).
Much more experience needs to be
gained before we can identify
safe situations where the generation
of 
``obviously correct'' and ``boring''
code might be considered a net
productivity boost.

Despite the negative tone,
many resounding formal methods successes exist: how the Javacard was formally verified~\cite{javacard-fv}, and how Intel's award-winning effort~\cite{roope-intel-fv-i7}  helped 
dispense with traditional testing in the design of floating-point hardware.
Over the decades, industrial corporations also have championed the advancement of formal methods around industrial products: Bell Labs in the 1980s and 1990s, Microsoft Research in later decades, and presently Amazon Web Services
which now employs a large number of scientists trained in formal methods, and employs formal methods heavily in-house~\cite{neha-billion-smt-queries}.
The spill-over benefits of this industrial know-how onto {\em openly accessible correctness research} are, barring a ready job market for students, indirect and scanty.

In this context, two additional
questions must be raised for which there seem to be no easy answers without additional research investments: (1)~for scientific computing software, there isn't a directly identifiable industry that will employ such in-house talent to check correctness, and (2)~with companies building their own hardware---such as Google building Tensor Processing Units, and Amazon, Microsoft, Tesla, and startups such as Groq building their own chipsets,
 it is likely that scientific computing software will begin using these specialized components---making the problem
of access to specifications even more challenging.
In the article by Reed et al. on HPC's future being cloudy and uncertain~\cite{cloudy-uncertain-reed}, this inevitability of using the hardware designed for AI is brought out as a necessity.


A 2017 DOE report on correctness in scientific computing~\cite{osti_1470989} 
exposed some of the emerging issues six years prior to the CSC 2023. 
Since then, the gap has continued to widen, as vividly captured in a recent survey~\cite{rising-heterogeneity} that focuses primarily on numerical issues but also
on surrounding issues pertaining to concurrency, compiler opacity, and rapid assimilation incorporating heterogeneity.
Added to this is the ability to bring high-performance computing to bear
on problems with an immediate horizon of relevance and deployment
such as the design of new 3D-printed rockets~\cite{relativity-space}, 
future power-grids~\cite{pnnl-energy-grid} and methods to 
target pandemics~\cite{covid-bell}.
%
%
While disasters in HPC\footnotemark~help promote the urgency of correctness methods,
the longer-term perspective of reliable science and productivity constitute the real priorities faced by the area.
\footnotetext{``Sometimes it takes an event like the crash of the Ariane 5 rocket, a naval propulsion failure or a crash in
a robotic car race to make people aware of the importance of handling exceptions correctly in numerical software''~\cite{DBLP:conf/correctness/DemmelDGHLLLPRR22}.}

In summary, renewed attention to {\em open} (non-industrial) correctness research is urgently needed because (1)~the diversity in the space of hardware and software has increased, (2)~there are too many undocumented components and APIs
and/or ones with shifting specifications, and (3)~without seeding an open research community focused on Correctness
for Scientific Computing, efforts ranging across technologies, tools, training and teaching can languish.
On a brighter note, the availability of
community resources such as
AI Benchmark Suites for Science~\cite{AI-Benchmarking-For-Science} are quite important for advancing correctness research in emerging areas of Scientific Computing.

\paragraph{Balancing Between Older Problems Versus Newer}
Given that we are playing 
catch-up with respect to
the number of unsolved correctness
challenges, we cannot
choose between long-lived codes
(e.g., for climate simulation)
and emerging codes/practices
(e.g., the use of 
language-level models).
Focusing on long-lived codes 
helps sustain them across newer
machines, languages and
compilers, while
focusing on emerging practices
can help forestall poorer
practices from taking root; hence
both are important to consider.
We now touch upon some
 important emerging challenges.

 The use of trustworthy ML-based surrogate models alongside HPC methods is 
 helping
 make advances in 
 many
 important HPC areas.
Unfortunately, these developments are 
happening when
we have not adequately
solved the reproducible inference behavior of textbook
neural networks such as LeNet~\cite{repro-nn-icse22}. 
In this work, 
a deep-learning (DL) model
is defined to be reproducible, if {\sl under the same
training setup (e.g., the same training code, the same environment,
and the same training dataset), the resulting trained DL model
yields the same results under the same evaluation criteria (e.g.,
the same evaluation metrics on the same testing dataset).}
This article points to the
randomness in the software (e.g., DL algorithms) and
non-determinism in the hardware (e.g., GPUs) as the main reasons for non-reproducibility.

Addressing reproducibility in this setting is a deep research challenge, ultimately requiring scalable formal semantics for the training processes, and also methods to attain reproducible behavior through the hardware/software stack of deep networks.
The title of a recent article
``The Grand Illusion: The Myth of Software Portability and Implications for ML Progress''~\cite{hooker-myth-2023-portability} pretty much says it all about the present state of reproducibility and portability:
{\em do no expect today's hardware/software ML stacks to offer these guarantees!}

 None of today's large-language models would have become a reality without the advent of GPUs,
without the use of mixed-precision training methods~\cite{pytorch-mpt}, and without the use of high levels of neural network sparsification.
Achieving impressive top-accuracy percentages is one metric of correctness; however, potential errors due to Pruning Identified Exemplars (PIEs~\cite{hooker-pie}) are important to include in  correctness discussions.

%

%

The use of low-precision floating point representations
comes with added risks 
such as increased chances 
of floating-point exceptions
or even show-stopper  
bugs during network training~\cite{nan-loss} where loss-gradient vectors uselessly collapse into vectors of NaNs~\cite{ieee-754-fp-standard}.
Once data values---even those that are not of direct interest externally---contain NaNs,
many errors are possible: (1)~the code becomes non-reproducible~\cite{demmel2022proposed} if the programmer does not correctly handle the logic of NaNs,\footnotemark and (2)~many control paths (even those that may contain unrelated serious bugs) become unobservable~\cite{demmel2022proposed}.
\footnotetext{A brain-teaser due to Prof. Nelson Beebe at Utah is to answer whether this macro is reliable when used for floating-point data: \#define MAX(x,y) ((x) >= (y) ? (x) : (y)). The answer is provided 
with the next footnote.}
These facts are highly disturbing
in the light of
research work~\cite{gpuverify:oopsla13} that has demonstrated that many widely used (and seemingly trusted) HPC and ML codes end up generating NaN floating-point exceptions---even on the very data-sets that were distributed  by the code-creators~\cite{hpdc2023-gpu-fpx}.
Added to this is the issue of
compiler-induced result variability~\cite{science-on-keel-cacm,DBLP:conf/sc/0007LR20,DBLP:conf/iiswc/SawayaBBGA17}
where the optimization flag settings
such as {\em fast math}
can visibly affect the behavior of the generated code.
Compiler-induced result variability
is slowly being recognized as a threat, but with no 
practical solutions
available at present. 
%
%
In fact,
not only do compiler flags change the floating-point results, they also can alter the exception-generation behavior of codes~\cite{hpdc2023-gpu-fpx}.
If these issues are not discussed in a timely manner, the reliability of scientific results~\cite{science-on-keel-cacm} may be gravely undermined in the era of rising heterogeneity.



  %


  The state of GPU weak memory-model definitions has advanced from~\cite{alglave-asplos15} to formal Alloy models~\cite{lustig-ptx-mem-model}.
%
Besides, there are now many GPU types~\cite{gpu-harbor} with subtle concurrency differences that are experimentally being discovered.

\paragraph{What is Not Detailed in this Report?}
We deliberately leave out a few 
topics mainly due to space limitations.
One such topic is achieving 
minimal acceptable performance.
This report sets aside
performance considerations
to help better focus on
behavioral correctness issues  that 
have not received nearly as much
attention as performance has.
Performance and correctness in fact go hand-in-hand
in any holistic approach (e.g., the Spectre bug mentioned earlier is an example where 
aiming for performance
led to incorrectness).
Another
is how a problem is modeled.
For example, considerations such as  how
the physics domain is
discretized 
are out of scope.

Despite our lack of 
elaboration, we 
reiterate that 
AI correctness in
general, 
and AI code generation
are important areas that will
have a huge impact on 
scientific computing correctness.
The mixed use of existing code 
and auto-generated code presents 
numerous correctness issues (e.g.,
how to maintain and port such codes)
that need {\em concerted} and
{\em proactive} action by the 
correctness
community.

\section{Details of
Areas, Techniques, Intervention Opportunities}
\label{sec:challenges}
We organize this section under the headings of 
Key Areas to Target (\S\ref{sec:key-areas-to-target}),
Key Techniques to Further (\S\ref{sec:key-techniques-to-further}),
and 
Key Intervention Opportunities
(\S\ref{sec:key-intervention-opportunities}).

\subsection{Key Areas to Target}
\label{sec:key-areas-to-target}
\subsubsection{Specification}

Formal specification is a prerequisite for verification: a verification tool needs to know how a program or hardware component is expected to behave before it can verify or refute that expectation.   Specification may also serve as a rigorous form of code documentation, and may be used in automated test generation.  A number of specification languages have been developed, targeting different programming languages and domains.  Examples include ACSL (ANSI/ISO C Specification Language) \cite{acsl-1.19}, used by the Frama-C analysis platform \cite{frama-c-27.1}; Verifiable C API specifications, part of the Verified Software Toolchain \cite{appel:2011:vst}; and SPARK contracts for Ada programs \cite{moy:2023:spark}.   These languages are typically used to specify pre- and post-conditions for functions, and various kinds of invariants.

Specification is difficult in any case, but scientific computing poses special challenges.  First is the \emph{oracle problem}: for many scientific applications, the correct result on inputs of interest is not known.  Furthermore, some applications compute approximate solutions to problems for which it is not even known whether smooth solutions exist \cite{fefferman:2022:navier-stokes}.

This should not be interpreted to mean the situation is hopeless.   An application is built from multiple components, many of which have clearly defined and well-understood responsibilities.   For example, an application may use a library for solving a system of linear or differential equations whose mathematical properties have been described and proved in the literature; or use numerical techniques which have been analyzed and have known properties of convergence and stability.  An application may use a message-passing library, such as MPI, which has a thorough, though informal, specification \cite{mpi40}.

These kinds of components are ripe for formal specification (and verification), but this will require specification languages to support appropriate abstractions for the domain of interest. For example, concepts like \emph{vector}, \emph{matrix}, \emph{matrix addition} and \emph{multiplication}, \emph{linear function}, \emph{function composition}, \emph{derivative}, and \emph{process group} could be realized as primitives in a language, which would make specifications much more easy to write, understand, maintain, and use in verification.

Even when an application or component cannot be specified completely, there are many opportunities for specifying particular correctness properties.  For example, many simulations are expected to conserve a physical quantity (e.g., energy) or particular symmetries; such properties can be specified as invariants.   Specifying metamorphic properties \cite{chen-etal:2018:metamorphic}, which assert a relation between outputs given a relation on the inputs 
(e.g., $a\leq c\wedge b\leq d \Rightarrow (a+b)\leq (c+d)$)
could also be extremely useful, especially when it is not possible to fully specify the output as a function of the input.

Finding appropriate ways to specify approximations is another challenge.  Many scientific applications are built upon layers of approximation, including:
\begin{enumerate}
\item iteration termination: using an iterative algorithm that converges to a solution only asymptotically and terminating it after a finite number of iterations;\footnotemark
\item discretization in time and space: approximating continuous functions by a discrete, finite set of points;
\item floating-point (round-off) error, for both IEEE arithmetic and newer systems; for computations which use variables with different precision; and for programs run on heterogeneous systems where different components use different floating-point systems;
\item stochastic error from Monte Carlo algorithms, discussed at significant length in~\cite{nelson-book}; and
\item lossy compression, used to reduce storage while still maintaining some specified accuracy guarantee.
\end{enumerate}
There is a need for effective ways to specify acceptable approximation error in each of these categories, and these specifications should be composable so that specifications of approximating components imply expected behavior of the whole application. 
\footnotetext{Answer to the brain-teaser: unreliable for Floating-Point because NaN compared to anything (including NaN) yields {\em false}. Thus if $x={\rm NaN}$, $y$ is returned and the user fails to see the NaN on $x$. Building on this, imagine the iterative algorithm ingesting a NaN in its loop termination predicate: it might exit permaturely or infinitely loop!}

An example of a desirable outcome along these lines was expressed by one participant as a wish for ``\ldots{}a solver for a relevant system of PDEs or ODEs, running on a parallel computer, with some guarantee that as long as the client satisfies some (checkable) conditions needed to guarantee a solution, the solver will return a solution to some guaranteed level of accuracy within some guaranteed amount of time.''   This goal requires a way to formally specify (1) preconditions that guarantee a solution, (2) postconditions that specify the accuracy, and (3) a way to specify performance (time).

Other specification issues include specification of memory models and instruction sets, again in a composable way; specification of compiler trans\-for\-mations---both so the user can specify when a transformation is acceptable, and for the compiler to report back to the user a description of what it has done; and the ability to write specifications for different programming languages that cooperate.
\subsubsection{Concurrency}

Much scientific software, and all HPC software, is parallel.  This poses a number of challenges.   First, much existing verification technology, especially that based on deductive verification, was designed for sequential programs.   There are some formal approaches that have exhibited moderate successes on concurrent programs; most notable are approaches based on model checking.  However, most of these tools target a particular concurrency language or model, e.g., C/C++ threads, or OpenMP, or MPI.    There is an urgent need for verification tools that can be applied to \emph{hybrid} concurrent programs that use a combination of these models.   This includes programs that use GPUs (or other devices) that have their own distinct concurrency models.   Yet another issue is the use of weak shared memory consistency models, which are extremely difficult to reason about informally, and have little support in current verification tools.  
\subsubsection{Abstractions and Approximations}
Modeling problems in the domain of scientific computing involves some
form of approximation.  For example, discretization and modeling
non-linear processes with linear systems is an example. Whether any
such algorithm converges to a solution (e.g., Navier-Stokes equations)
is not the focus of the exploration. Checking the correctness of such
discretization schemes needs the involvement of the domain scientist.
Given such discretization schemes, there are numerous other
approximations that are carried out in various parts of the HPC
computing stack which can involve computation with finite precision,
compression, surrogate models, and other mathematical approximations.


\subsubsection{Numerical Challenges}
The eventual computation of the various algorithms is performed in a
finite precision representation that is natively supported in
hardware. Any finite precision representation has limitations on the
dynamic range (i.e., range of values that can be represented) and the
precision (i.e., the accuracy with which each value is represented).
The single and double precision formats in the floating point~(FP)
representation is widely used in scientific computation.

Many applications are essentially enabled by high performance beyond
the performance of single precision and double precision formats.  For
example, the deep learning revolution that started only after GPUs
became widely available. This also applies to scientific applications;
for example, one of the biggest break-through in improving the
accuracy of climate simulations would be to enable kilometer-scale resolution
- a challenge that requires high-performance computing and data
management. 
In support of these heightened computational demands,
numerous representations that trade-off the
accuracy for more performance are being developed, but currently without proper support
for overall correctness.

Low precision representations that trade-off the dynamic range and
precision are becoming increasingly common in the ML domain (e.g.,
\texttt{Bfloat16}~\cite{DBLP:conf/date/TagliaviniMRMB18},
Posit~\cite{DBLP:journals/superfri/GustafsonY17},
and \texttt{TensorFloat32}~\cite{nvidia:tensorfloat:online:2020}).
The correctness of systems and the interpretability of results with
the use of low precision
and mixed-precision representations is a significant
challenge~\cite{various-precision,ichi-gmres}.
It requires new mathematical libraries~\cite{cgal},
error
analysis
algorithms~\cite{fptaylor,satire,aanjaneya:rlibm-prog:pldi:2022,lim:rlibm32:pldi:2021,lim:rlibmall:popl:2022,lim:rlibm:popl:2021},
static and dynamic analysis tools to explore the trade-off between accuracy and performance~\cite{DBLP:conf/sc/Rubio-GonzalezNNDKSBIH13,DBLP:conf/icse/Rubio-Gonzalez016,DBLP:conf/popl/ChiangBBSGR17, DBLP:conf/issta/GuoR18,Wang-icse24,DBLP:conf/supercomputer/LagunaWSB19, DBLP:conf/iccps/DarulovaHS18} and to reason about the
errors incurred~\cite{Chowdhary:positdebug:2020:pldi,Chowdhary:pfpsanitizer:FSE:2021,Chowdhary:EFTSanitizer:OOPSLA:2022,DBLP:conf/kbse/FrancoGR17}.
As numerical behavior is a function of input
ranges, proper input generation is also
an increasing challenge~\cite{dolores-ceil,parco-xscope,DBLP:conf/ppopp/ChiangGRS14,DBLP:conf/icse/0007R20,DBLP:conf/pldi/FuS17,DBLP:conf/popl/BarrVLS13,DBLP:conf/apsec/YiCMJ17}.
  

\subsubsection{Compression}

With the exponential growth of data, efficient data types and
compression algorithms have emerged to tackle the storage and
processing challenges. These advanced data representations must adhere to
certain criteria to ensure their effectiveness:

\begin{itemize}
\item They should be ``correct'' meaning they should avoid introducing noise that skews 
the behavior of downstream processes
while faithfully
representing the input values.

\item They should not alter the scientific analysis for which the data is used, ensuring that the
integrity of the analysis is maintained.


\end{itemize}  

However, despite their significance, there need to be more tools
available to analyze the impact of these new data representations in
advance. This creates a situation where users must rely on guesswork
to determine the most suitable method for their needs. The absence of
comprehensive analysis tools challenges
the widespread adoption and optimization of these data types.

In parallel to these advancements, computing systems have evolved
towards greater heterogeneity, deviating from traditional
HPC systems. This shift introduces
challenges related to system reliability. To address these challenges,
algorithms must be adjusted to accommodate the heterogeneity and
potential unreliabilities of the system. This involves leveraging the
inherent properties of the algorithms to account for device failures,
data corruption, and communication bottlenecks. The impact of
unreliabilities can be mitigated by designing fault-tolerant
algorithms that incorporate error-checking mechanisms, redundancy, and
intelligent load balancing.

\subsubsection{Surrogate Models and Mathematical Approximation}
Surrogate models, where one approximates a complex process with a
mathematical approximation, is becoming more prevalent in scientific
computing. Further, the use of machine learning methods as surrogate
models is becoming more common.  For example, some approaches that
approximate physics-based simulation with data-driven methods with a
neural network are common in robotics. Training these surrogate models
often requires
detailed
simulations that can
be expensive.
Reducing this cost while
obtaining acceptably
high training  accuracy
remains a challenge.




\subsubsection{Heterogeneity}
As Moore's law becomes increasingly expensive to sustain, novel architectures (conventional accelerators and quantum computers) are an important opportunity to continue deliver greater computing performance and efficiency.
Hardware accelerators, e.g., GPUs, field programmable gate arrays (FPGAs), and application specific integrated circuits (ASICs), have for the past dozen years been seen as the way to use abundant transistors with tight power constraints.
A disappointing reality has been that mapping broad classes of workloads to a few hardware designs for FPGAs and ASICs has been a difficult search process with limited clear examples of success—hardware acceleration has been commercially successful in cryptography, media codecs, and network protocols, but outside of those well defined problems, it has been hard for specialized hardware to flexibly support broad classes of problems.
Scientific computing is one area where innovation in mapping problems to hardware can succeed: Scientific computing problems (PDEs, optimizations) have well-standardized problem statements, with richly interrelated solution algorithms to solve problems.
While there have been many user-friendly examples of scientific computing taking advantage of GPUs, there has been much less infrastructure available for scientific computing to make use of configurable or custom hardware.
Research that aims
to fill this void can 
potentially bring
about significant
 cost/energy savings 
in applicable domains.


\subsubsection{Interoperability}
\label{subsec:interop}

Scientific workflows increasingly dominate the job mix of distributed HPC systems.\footnote{\url{https://workflows.community/systems}.} Data acquisition, processing, analysis, and dissemination through distributed network workflows e.g. \cite{htcondor,pegasus2019,panda2014} are central to scientific discovery in diverse domains including astronomy \cite{lsst}, particle physics \cite{lhcpanda}, bioinformatics \cite{ampl}, climate modeling \cite{gmd-13-5567-2020}, among many.
Drivers include the integration of simulation with analysis, requiring a collection of interconnected tasks often bridging multiple data sources and networks.  Workflows encompass ensemble jobs in a single cluster in which results from a set of runs determine the path through a task graph. Workflows can span multiple computing sites, with  simulation and analysis tasks being augmented with data management, provenance tracking, and storage of data sets and results.

While a variety of workflow management systems exist, it is challenging for 
HPC and scientific computing programmers to adopt and
integrate correctness-checking practices into their workflows. There are multiple factors affecting the correctness of a workflow. Often workflow tasks are written in different languages such as C++, Fortran, Python and other scripting languages such as shell scripts, or domain-specific frameworks. The tasks may incorrectly assume the locations of data sets, encounter permission issues, or even experience errors in the distributed messaging layers. Many of these problems are not related to the correctness of the actual computational task, but lack of exception handling may lead to errors (potentially undetected) that compromise the validity of the results.
\subsection{Key Techniques to Advance}
\label{sec:key-techniques-to-further}



\subsubsection{Deductive Verification}
The most customizable and rigorous approaches to program verification use automated or semi-automated theorem provers, including Satisfiability Modulo Theories solvers and interactive proof assistants, to establish correctness properties grounded in foundational logic and formal semantics. This is very appealing as a way to mathematically guarantee correctness.
 
For digital systems in general, such deductive verification is powerful but typically requires considerable investment of time and expertise. For scientific computing, particular challenges of deductive verification arise from the complexity in many aspects:
\begin{itemize}
\item Correctness properties that are highly application-specific
\item Numerical approximation issues (iterative convergence, floating point, etc.) that complicate the exact definition of correctness
\item Large scales of parallelism that increase potential nondeterminism in behavior (both ``ordinary'' such as execution interleavings, and ``extraordinary'' such as hardware faults)
\item Large existing code bases (including libraries) that are not formally specified or verified
\item Numerous software-stack layers that underpin scientific applications
\item High-performance languages and compilers that may not conform to rigorous semantics
\end{itemize}
 
Demonstrations of end-to-end modular foundational library/tool-based machine-checked verifications of numerical software are already appearing in the formal methods literature over the past 5 to 7 years (e.g., \cite{2023-arith, tekriwal-etal:2023:jacobi}) --- but they are small-scale; now it's time to scale them up.

Modular and incremental verification approaches are especially important because, while the ideal is a complete end-to-end proof, practical work initially targets portions of the abstraction hierarchy and this can provide much insight. For example, developing candidate formal specifications (for existing or new code modules) can clarify the intended behavior and help detect or prevent bugs even if formal verification is not yet carried through to the implementation. Likewise, discovering useful invariants satisfied by existing code can provide confidence and support further reasoning even short of a full correctness proof.
 
Academic tools for deductive verification are not currently usable on large, parallel code bases. Even those that do exist support only subsets of the desired feature set. One question we are faced with is how to integrate formal methods tools seamlessly into workflows. The opportunity here is great, since even the process of writing some formal semantics reveals potential optimizations and errors because it mechanizes deep knowledge of the systems being developed.

Lowering the barrier to incremental use of deductive verification for scientific computing will promote the use of these formal approaches starting early in the development process, where they have the best chance of preventing defects and producing code that can be tractably proved correct because it is ``designed for verification''. To this end, there is value in creating and extending formal domain-specific languages and libraries for specifications and proofs.
The application domain for 
these languages/libraries 
might be
HPC as well as AI/ML.
%

\subsubsection{Model Checking}
Model checking encompasses a variety of techniques for verifying correctness properties of a program, protocol, or concurrent system.  These techniques typically involve (1) expressing the system of interest as a finite-state transition system in some way, (2) expressing the properties of interest in a formal way, such as temporal logic or assertions, and (3) using a "brute-force" automated algorithm to enumerate the states of the system and check the properties.  

There have been a number of successful small applications of model checking to scientific computing, see, e.g., CIVL
\cite{siegel-etal:2015:civl_sc}.   These approaches require the user to specify a (typically small) bound on the sizes of input data structures, number of threads or processes, and other parameters.   The bounds enable a tool to automatically construct a finite-state model of the program, which can be explored in full detail.   This approach has been used to verify a Conjugate Gradient code \cite{huckelheim-etal:2018:sas}, various matrix multiplication routine (including an implementation of Strassen's algorithm \cite{siegel-verifythis:2016}), a parallel Gaussian Elimination code, and finite difference code generated by Devito \cite{huckelheim-etal:2017:correctness}, among many other examples.
Developing methods to compute these bounds as well as determining how small these
bounds can be made are open questions
that can significantly benefit from case studies.

\subsubsection{Static Analysis}
Static analysis methods 
for checking generic properties:
absence of data races, deadlocks, 
and memory errors are highly desirable.
Methods that target these bugs while
statically analyzing MPI, OpenMP and GPU programs
will be very popular, but requires concerted 
community-wide effort to develop---for instance to determine properties amenable to static analysis, 
gauge the level of false-positive rates, and finally to understand
the effort to build suitable
infrastructures.
Such analyzers must ideally generate
very few false alarms, but also detect sufficiently many true bugs.

\subsubsection{Fuzz Testing}
Fuzz-testing~\cite{miller-fuzzing} has 
enjoyed significant industrial
adoption, but is yet to gain
 traction in testing scientific
 programs.
One reason might be the fact that the 
input constraints are typically not expressed for many HPC program---or this specification may be too nuanced, and hence stated only loosely at a very high level.
This makes it impossible to know whether one is fuzzing within legitimate input ranges.
However, given the success of coverage-guided fuzzing in other areas, it is important to introduce fuzzing into HPC---perhaps targeting input values more than control-flow paths.

\subsubsection{Debugging}
Given that 
root-causing bugs is as important
as triggering it in the first place,
one must focus on this problem.
Typical focal points 
 might include the creation of
 minimal bug instances through
  delta-debugging which has enjoyed some success in HPC testing to date~\cite{flit-hpdc19}.
 Tools such as 
 CSmith~\cite{csmith} are important to transplant into the HPC domain as well (a good start in this direction is Varity~\cite{varity}).

\subsection{Key Intervention Opportunities}
\label{sec:key-intervention-opportunities}
These are the elements to which the fundamental techniques
and tools will be applied.

\subsubsection{Scientific Workflows}
As discussed in Section \ref{subsec:interop}, distributed workflows spanning HPC clusters, experimental facilities, and storage systems have become ubiquitous in scientific discovery. Verifying correctness of workflows affects a multiplicity of components, often dynamically assembled. Challenges include optimizing performance while validating correctness in orchestrating and combining tasks and their dependencies across multiple programming languages and communication protocols. The diverse nature of workflow components require a full suite of validation tools. Components separately developed in Python, Go , C$++$, Fortran, and domain-specific languages must be composed to execute on diverse hardware. Distributed computing protocols are employed to communicate events and data streams comprising a workflow. While each of these separately present severe correctness analysis challenges, errors due to concurrency, non-determinisim, and resource constraints pose higher levels of difficulty in validation. Anecdotally, workshop participants discussed debugging an intermittent failure in a ML workload in a cloud environment. With considerable difficulty, the root cause was isolated to non-determinism of message transmission leading to timeouts and buffer insufficiency.
While the DOE ASCR program recently funded several research efforts to analyze and optimize performance of extreme scale workflows, systematic application of correctness tools to scientific workflows is lacking. 

\subsubsection{Application}

An application is an executable set of software components and general-purpose libraries that have been harnessed together to solve a specific class of problems in a scientific domain.   Examples of applications include FLASH (astrophysics) , OpenMC (neutron transport), and EQSIM (earthquake simulation).  Ideally, an application will make extensive use of libraries and other reusable software components that have been fully specified and verified.  This will simplify the specification and verification of the application, which should require only the contracts, and not the implementation details, of its components.  Also ideally, an application will have a modular design, with clearly defined interfaces, so that each module can be specified and verified individually.

\subsubsection{Verified Libraries}

Creating verified  libraries
is a direction that can have
significant benefits. The possibilities in this space are briefly listed under ``challenge problems''
on Page~\pageref{page-of-challenges}.

\subsubsection{Program Synthesis}
In its most generic definition, program synthesis may be considered a class of techniques that can generate a program from  semantic intent. Examples of program synthesis techniques include specification languages, declarative programming, automatic differentiation, symbol regression and most recently Large Language Models (LLM). 




LLM-based code generation has attained significant popularity and acceptance. In HPC, driven by hardware heterogeneity, LLMs have been employed to generate scientific software that uses portable abstraction libraries such as RAJA and Kokkos. Manual code development becomes prohibitive as heterogeneous devices proliferate, yet reliance on LLM and other automatic techniques poses new challenges for verifying correctness of generated code. While LLMs are trained on an enormous collection of code repositories and discussion forums, existing code and discussion out ``in the wild'' can be error-prone and misleading. Establishing correctness of training data with correctness tools would improve robustness of the model. Expressing intent in natural language may be ambiguous and imprecise, leading to multiple and potentially conflicting interpretations realized in code. Generating a collection of programs for the very same query and comparing their execution results through rigorous unit testing is  a more systematic approach, as is formal verification of the generated programs for equivalence.

\subsubsection{Programming Languages}

Programming languages can play an important role in ensuring correctness.   For example, static type checking, enforced by the language and compiler, prevents the programmer from committing innumerable errors.  We expect further developments to provide  stronger correctness guarantees.   Novel language approaches for expressing parallelism, for example, can prevent nondeterminism bugs, data races, and deadlocks~\cite{vivek-sarkar-async-finish}.  Domain-specific languages, such as ATL (A Tensor Language~\cite{atl-language}), and high-level languages designed for scientific computing, such as Julia, may make it easier to verify correctness by enabling reasoning at a higher level of abstraction than that of languages like C or C++.

\subsubsection{Verified Compilers and Build Systems}


Compiler verification in HPC
has to go beyond  the conventional bit-equivalence preservation in order to support interesting 
optimizations.
One research question to ponder is whether compiler optimization flags must carry a range of allowed semantics.
This may mean allowing various levels of departure---all of which are formally analyzable---from an established baseline semantics.
If an HPC-oriented compiler provides a way to recover deterministic semantics, the fact that it does so will be an interesting and (relatively) well-defined goal.

\section{Concluding Remarks}
\label{sec:concl}
The DOE/NSF Workshop on Correctness in Scientific Computing (CSC)
held on June 17, 2023, as part of the Federated Computing Research Conference
(FCRC) was in response to
concerns of a growing trust-gap
if we let the hardware/software complexity along with scaling ambitions
of HPC   go unfettered without a timely well-rounded
discussion around correctness.

The workshop attracted experts from academia and laboratories.
The absence of participation from
participants who have
direct industrial appointments was 
compensated
to a large measure
by industrial perspectives that the participants nevertheless possessed. 
The areas represented included
formal specifications,
concurrency, approximations,
heterogeneity,
interoperability,
and operational challenges.

Several opportunities and directions were pointed out, especially by the invited speakers:

\begin{itemize}
    \item 
How do we provide end-to-end mechanically-checked guarantees for computational algorithms and implementations,  generalizing
them to be widely applicable?

\item Can we extract mechanically-checked bound information on error? How useful are these formal bounds? Must we develop useful probabilistic bounds?

\item Verifying correctness in new numerical formats is crucially important.

\item   Assertions that detect error early on
are important. For 
example, 
while pushing
code into repositories,   
many correctness
checks could potentially be 
performed.

\item While ML in scientific computing
is important, when ML models are
interfaced with HPC, we must understand how error builds up.

\item Compiler optimization guarantees
and the 
verification of runtimes and libraries
are important.

\item Verification methods must be
integrated within HPC languages and intermediate representations such as LLVM and MLIR.

\end{itemize}

\subsection{Outreach and Education}

In order to convey correctness concepts
in a tangible and HPC-relevant manner,
formal tools must be situated around
languages and formalisms already familiar
to HPC practitioners.
Languages such as 
the Concurrency Intermediate Verification Language (CIVL~\cite{siegel-etal:2015:civl_sc})  and associated
verifiers do support this practice.
It sems logical that they can be
harnessed to 
teach the virtues of
following widely recommended
concurrency patterns~\cite{wenmei,dpcpp}.
Also their 
verifiers may also permit formal methods
competitions to be precisely defined and conducted.
Benchmark suites such as the Indigo Suite that
contain a calibrated set of GPU Irregular Codes with options to enable data races~\cite{indigo}
and DataRace Bench~\cite{datarace-bench}
designed for OpenMP are excellent starts---but clearly, many more such benchmark suites are essential, given the rising heterogeneity.
As to floating-point, resource 
pages such as FPBench~\cite{fpbench} are good starts, but in need for significant expansion covering modern heterogeneous architectures and floating-point standards.

Successful outreach
requires
steadfast
commitment toward
organizing and running
regular meetings,
moderating discussions,
and maintaining 
accurate and
consistently updated
websites.
One example of such an
outreach approach is the
monthly meetings organized  by the FPBench group (\url{https://fpbench.org/}) that brings together worldwide talent to discuss numerical correctness checking and correctness tooling issues.
They also conduct
a yearly workshop called ``FPTalks'' 
that creates an incentive for researchers at all stages to field their ideas.

Successful outreach also
requires one not to
wastefully
duplicate effort.
For instance, the IEEE Technical Community
on Parallel Programming
(\url{https://tcpp.cs.gsu.edu/curriculum/}) has had an outstanding track record of
promoting education in parallelism and concurrency.
It may be worth capitalizing on
their established status and seek cooperation before launching another 
similar effort.
This can help better publicize concurrency verification workshops and ``bug-fests'' events.


Two-way dialog involving 
  HPC researchers (who may be unfamiliar
  with mature formal methods)
  and  formal methods 
  researchers whose education
  is largely based on today's trendy
  computer science classes
  (who may be unfamiliar with
  linear algebra) also seems essential.
To illustrate this need, we put forth
two challenges that need concerted effort to solve:
\begin{itemize}
\item 
Consider the problem of understanding
the impact of
rounding errors
on the condition number of a matrix.
Specifically, consider a matrix with a known condition number (in the real-number
domain)  being instantiated in limited
precision and operated upon by today's
(often IEEE-noncompliant) tensor processing
units.
Does it matter how we round, or must one 
build in precautions such as numerical
correction?

\item  Adoption of languages that
emphasize type-based memory
safety
(e.g., Rust~\cite{rust})
into the world of GPU programming
is in its 
infancy~\cite{rust-cuda,embark-studios-rust-gpu}. Even if programmers employ
such Rust dialects without using 
too many ``{\tt unsafe}'' annotations
on realistic GPU codes,
the extra burden of incorporating Rust-programming methods  needs to be assessed.

\end{itemize}

\subsection{Tool Concurrency, Interoperability, Libraries}

A pain-point will likely become
apparent if one tries to apply today's
verification methods at scale: that
the verification tools themselves
are ill-equipped to exploit 
parallelism and concurrency.
  The availability of parallelism in an HPC system can be taken as an incentive for traditional correctness tools   to be parallelized.

 With increasing emphasis on interoperability
 and with increasing
 use of workflows, 
 it becomes important to
  develop declarative methods to specify the overall computation to be achieved,
  while performing associated
  safety checks.
  These checks
may include those
that ensure that
the computation is working on the right versions of the data, and abides by the 
 declared permissions.

Given the large number of
libraries and the need for creating more of them, perhaps differential testing methods to compare new libraries against established ones---or check the behavior of a new functionality against a composition of established (and trusted) functions---appears as an attractive avenue to pursue.
This may handle the ``oracle problem'' in
many situations, because today's libraries
often have multiple implementations.

It is also important to consider the extent to which a library
abstraction is preserved when its
implementations are migrated to take advantage
of emerging heterogeneous architectures.
Do such migrations preserve the overall
library semantics, or would it be necessary
to export semantic adjustments
to its interface specifications?

\subsection{Multiple Programming Languages and Scientific Domains}

In the modern context, scientific computing
depends on multiple programming languages: Fortran continues to be important, Python  readily offers a path to ML and  domain-specific languages, with the dominant emerging pragmatic choice for performance often being C++ and its dialects.
Languages such as Julia~\cite{churavy2022bridging} have a good following, as it offers flexibility
while taking advantage of  modern programming language technologies
at all levels (multiple dispatch, LLVM
integration, even smooth use of GPUs and other accelerators).
The choice of programming languages impacts correctness definitions and verification methods, with many avenues still unexplored.


More than just the language and its 
compiler,
what is lacking at the source code level is what abstraction is being realized.
These may be from thousands of domains (chemistry, physics, biology, cosmology, statistics, etc.) and computational spaces  (derivatives as in automatic differentiation, frequencies in Fourier spaces, etc.). 
Preserving the higher level domain/space semantics through compilation, optimization, and hardware implementation---when these lower-level entities are often designed without awareness of what is tolerable at the higher levels---remains a central challenge in scientific computing.

\clearpage 

\appendix
\section{Appendices}

\subsection{Workshop Program and Participants}
\label{sec:csc-program-gist}

The CSC program consisted of two keynotes, lightning talks, and breakout and readout sessions:
\vspace{.75ex}

{
  \newcommand{\twidth}{.5\textwidth}
  \newcommand{\myskip}{3.65   ex}
  \begin{tabular}{ll}
    \toprule
    Keynote Speakers
    &
    \\
    \midrule
    \begin{tabular}[t]{l}
      Ignacio Laguna\\
      Lawrence Livermore National Laboratory
    \end{tabular}
    & \parbox[t]{\twidth}{\it Letting HPC Programmers Focus
      On Correctness First, Then On Performance}
    \\[\myskip]
    \begin{tabular}[t]{l}
      Jean-Baptiste Jeannin\\
      University of Michigan at Ann Arbor
    \end{tabular}
    & \parbox[t]{\twidth}{\it Formal Verification
      in Scientific Computing}
    \\[\myskip]
    \midrule
    Lightning Talk Speakers
    &
    \\
    \midrule
    \begin{tabular}[t]{l}
      Aditya V.\ Thakur\\
      University of California, Davis
    \end{tabular}
    & \parbox[t]{\twidth}{\it Towards Provable Neurosymbolic
      Training of Deep Neural Networks for Scientific Machine Learning}
    \\[\myskip]
    \begin{tabular}[t]{l}
      Alyson Fox\\
      Lawrence Livermore National Laboratory
    \end{tabular}
    & \parbox[t]{\twidth}{\it Analytical Bias in ZFP Lossy
      Compression for Floating-Point Data}
    \\[\myskip]
    \begin{tabular}[t]{l}
      Andrew W.\ Appel\\
      Princeton University
    \end{tabular}
    & \parbox[t]{\twidth}{\it Formally Verified Numerical Methods}
    \\[\myskip]
    \begin{tabular}[t]{l}
      Andrew Siegel\\
      Argonne National Laboratory
    \end{tabular}
    & \parbox[t]{\twidth}{\it Developing HPC Applications at
      the Forefront}
    \\[\myskip]
    \begin{tabular}[t]{l}
      Harshitha Menon\\
      Lawrence Livermore National Lab
    \end{tabular}
    & \parbox[t]{\twidth}{\it Ensuring Correctness in Programs
      Generated by LLM: Challenges and Solutions}
    \\[\myskip]
    \begin{tabular}[t]{l}
      Laura Titolo\\
      NIA/NASA LaRC
    \end{tabular}
    & \parbox[t]{\twidth}{\it ReFlow: from real number
      specifications to floating-point number implementations}
    \\[\myskip]
    \begin{tabular}[t]{l}
      Piotr Luszczek\\
      University of Tennessee, Knoxville
    \end{tabular}
    & \parbox[t]{\twidth}{\it Exception Handling in Programming
      Languages and Numerical Libraries}
    \\[\myskip]
    \begin{tabular}[t]{l}
      Sreepathi Pai\\
      University of Rochester
    \end{tabular}
    & \parbox[t]{\twidth}{\it Formal ISA Semantics Can Bind Us
      All (in the Quest for Correctness)}
    \\[\myskip]
    \begin{tabular}[t]{l}
      Vivek Sarkar\\
      Rice University
    \end{tabular}
    & \parbox[t]{\twidth}{\it Vivek Sarkar: Determinacy =
      Functional Determinism + Structural Determinism}
    \\[\myskip]
    \midrule
    Breakout Session Topics &\\
    \midrule
    \multicolumn{2}{@{\hspace{2ex}}l}{Formal Methods, Concurrency} \\
    \multicolumn{2}{@{\hspace{2ex}}l}{Number Systems, Precision} \\
    \multicolumn{2}{@{\hspace{2ex}}l}{Accelerators, Extreme Heterogeneity} \\
    \multicolumn{2}{@{\hspace{2ex}}l}{Future Compilers, Correctness Issues} \\
    \multicolumn{2}{@{\hspace{2ex}}l}{High Productivity Programming Languages,
    Static Analysis} \\
    \multicolumn{2}{@{\hspace{2ex}}l}{Productivity, Libraries, Sustainability}\\
    \multicolumn{2}{@{\hspace{2ex}}l}{Fault Tolerance and Resilience}\\
    \multicolumn{2}{@{\hspace{2ex}}l}{Correctness Challenge Problems for HPC}\\
    \bottomrule
  \end{tabular}
}
\clearpage

\subsection{Attendee List}
 
\begin{small}

\begin{longtable}{|l|l|r|}
        \hline
        First Name & Last  & Affiliation   \\
        \hline
Mridul &  Aanjaneya &      Rutgers University   \\
Andrew & Appel &      Princeton University  \\
Thomas &   Applencourt &      Argonne National Laboratory     \\
Dorian &   Arnold &      Emory University   \\
Alexander &   Bagnall &      Ohio University\\
Anindya &   Banerjee &      National Science Foundation   \\
David &   Bindel &      Cornell University\\
Franck &   Cappello &     Argonne National Laboratory\\
Michael &   Carbin &      Massachusetts Inst. of Technology\\
Florina &   Ciorba &      University of Basel      \\
Myra &   Cohen &      Iowa State University\\
Dilma &   Da  Silva  &    National Science Foundation   \\
Damian & Dechev & National Science Foundation \\
Peter &   Dinda &      Northwestern University\\
Irina & Dolinskaya  & National Science Foundation \\
Anshu &   Dubey &      Argonne National Laboratory\\
Karthikeyan &   Duraisamy &      University of Michigan\\
Hal &   Finkel &      Department of Energy   \\
Alyson &   Fox & Lawrence Livermore National Laboratory\\
Maya &   Gokhale &      Lawrence Livermore National Laboratory\\
Ganesh &   Gopalakrishnan &      University of Utah\\
Mary &   Hall &      University of Utah\\
Jeffrey &   Hittinger &      Lawrence Livermore National Laboratory\\
Torsten &   Hoefler &      ETH Zürich\\
Yipeng &   Huang &      Rutgers University\\
Costin &   Iancu &      Lawrence Berkeley National Laboratory\\
Jean-Baptiste &   Jeannin &      University of Michigan\\
Ignacio &   Laguna &      Lawrence Livermore National Laboratory\\
Ang &   Li &      Pacific Northwest National Laboratory\\
Piotr &   Luszczek & 
Innovative Computing Laboratory, University of Tennessee\\
Harshitha &   Menon &  Lawrence Livermore National Laboratory\\
Prabhat &   Mishra &      University of Florida\\
Jackson &   Mayo &      Sandia National Laboratories\\
Vijay &   Nagarajan &      University of Utah\\
Santosh &   Nagarakatte &      Rutgers University\\
Sreepathi &   Pai & University of Rochester\\
Sam &   Pollard &      Sandia National Laboratories\\
Jonathan &   Ragan-Kelley &      Massachusetts Inst. of Technology\\
Cindy &   Rubio-Gonzalez &      University of California Davis\\
Seungjoon & Park & National Science Foundation \\

\hline  

\end{longtable}

  \clearpage
 
{\hfill {Attendee List (continued)} \hfill}

\begin{longtable}{|l|l|r|}
         \hline
        First Name & Last  & Affiliation   \\
        \hline
Vivek &   Sarkar & Georgia Inst. of Technology\\
Markus &   Schordan &      Google\\ 
Andrew &   Siegel &      Argonne National Laboratory\\ 
Stephen &   Siegel &      University of Delaware\\
 Scott &   Stoller &      Stony Brook University\\
Zachary &   Tatlock &      University of Washington\\
Michela &   Taufer &      University of Tennessee Knoxville\\
Keita &   Teranishi &      Oak Ridge National Laboratory \\
Aditya &   Thakur &      University of California Davis  \\
Heidi &   Thornquist &      Sandia National Laboratories\\
Laura &   Titolo &    National Aeronautics and Space Administration\\
        \hline       
\end{longtable}    
  
\end{small}

\clearpage

\bibliographystyle{acm}
\bibliography{HPC-Correctness}
\end{document}